\documentclass
[aps,prl,twocolumn,floatfix,english,showpacs,10pt,superscriptaddress]{revtex4-2}
\usepackage{graphicx}
\usepackage{booktabs}
\usepackage{amsmath}
\usepackage{physics}
\usepackage{amssymb}
\usepackage{colordvi}
\usepackage{verbatim}
\usepackage{xcolor}
\usepackage{mathrsfs}
\usepackage{epsfig}
\usepackage{lipsum}
\usepackage{amsfonts}
\usepackage{makecell}
\usepackage[unicode=true, breaklinks=false, pdfborder={0 0 1}, backref=false,
colorlinks=true, linkcolor=blue, urlcolor=blue, citecolor=blue]{hyperref}%
\providecommand{\U}[1]{\protect\rule{.1in}{.1in}}
\setcitestyle{numbers,square}

\begin{document}

\title{Chiral-Damping--Enhanced Magnon Transmission}

\author{Xiyin Ye}
\affiliation{School of Physics, Huazhong University of Science and Technology, Wuhan 430074, China}

\author{Ke Xia} 
\affiliation{School of Physics, Southeast University, Jiangsu 211189, China}

\author{Gerrit E. W. Bauer}
\affiliation{WPI-AIMR and Institute for Materials Research and CSIS, Tohoku University, Sendai 980-8577, Japan}
\affiliation{Kavli Institute for Theoretical Sciences, University of the Chinese Academy of Sciences, Beijing 100190, China}
 
\author{Tao Yu}
\email{taoyuphy@hust.edu.cn}
\affiliation{School of Physics, Huazhong University of Science and Technology, Wuhan 430074, China}

\date{\today }

\begin{abstract}
The inevitable Gilbert damping in magnetization dynamics is usually regarded as detrimental to spin transport. Here we apply a general feature of chiral non-Hermitian dynamics to a ferromagnetic-insulator--normal-metal heterostructure to show that the strong momentum dependence and chirality of the eddy-current-induced damping also causes beneficial scattering properties: A potential barrier that reflects magnon wave packets becomes unidirectionally transparent in the presence of a metallic cap layer. Passive magnon gates that turn presumably harmful dissipation into useful functionalities should be useful for future quantum magnonic devices.
\end{abstract}
\maketitle

\textit{Introduction}.---Magnonic devices save power by exploiting the collective excitations of the magnetic order, i.e., spin waves or their quanta, magnons,  for non-reciprocal communication, reprogrammable logics, and non-volatile memory functionalities~\cite{Lenk,Chumak,Grundler,Demidov,Brataas,Barman,Yu_chirality,non_Hermitian_magnons,magnon_memory,Flebus}. The possibility to modulate magnon states and their transport in ferromagnets by normal metals or superconductors brings functionalities to spintronics~\cite{Bergeret,study_1,study_2,Weihan}, quantum information~\cite{Q_information,Q_information_1,Q_information_2,Q_information_3,Q_information_6,Q_information_4,Q_information_5}, and topological materials~\cite{topological,topological_2,topological_3,topological_4,topological_5,topological_6}.  The prediction of inductive magnon frequency shifts by superconducting gates on magnetic insulators~\cite{superconductor_gating_theory,similar_theory,superconductor_gate,superconductor_gate_1,superconductor_gate_2,superconductor_gate_3,superconductor_gate_4} have been experimentally confirmed~\cite{Borst}. Normal metals are not equally efficient in gating magnons~\cite{Ohmic_conductors2,Ohmic_conductors3,Ohmic_conductors4,Ohmic_conductors5}, but the stray fields of magnetically driven ``eddy currents"~\cite{eddy_currents1,eddy_currents2,eddy_currents3,eddy_currents4,eddy_currents5,eddy_currents6,eddy_currents7,eddy_currents8,eddy_currents9,eddy_currents10,eddy_currents11,eddy_currents12,eddy_currents13,eddy_currents14,eddy_currents15}  significantly brake the magnetization dynamics~\cite{eddy_currents1}.

The intrinsic Gilbert damping seems to be detrimental to transport since it suppresses the magnon propagation length. However, in high-quality magnets such as yttrium iron garnet (YIG) films, this is not such an issue since the magnon mobility is often limited by other scattering processes such as two-magnon scattering by disorder, and measurements can be carried out in far smaller length scales.  

Natural and artificial potential barriers are important instruments in electronics and magnonics by confining and controlling the information carriers. They may guide magnon transport~\cite{Borst,barrier_1}, act as magnonic logic gate~\cite{barrier_2}, induce magnon entanglement~\cite{semitransparent,Q_information_3}, and help detecting exotic magnon properties~\cite{barrier_3,barrier_4,barrier_5,barrier_6}. In the linear transport regime, the transmission of electrons and magnons through an obstacle has always been assumed to be symmetric, i.e., the same for a wave or particle coming from either side.

\begin{figure}[htp]
\centering
\includegraphics[width=0.42\textwidth,trim=0.4cm 0cm 0cm 0.1cm]{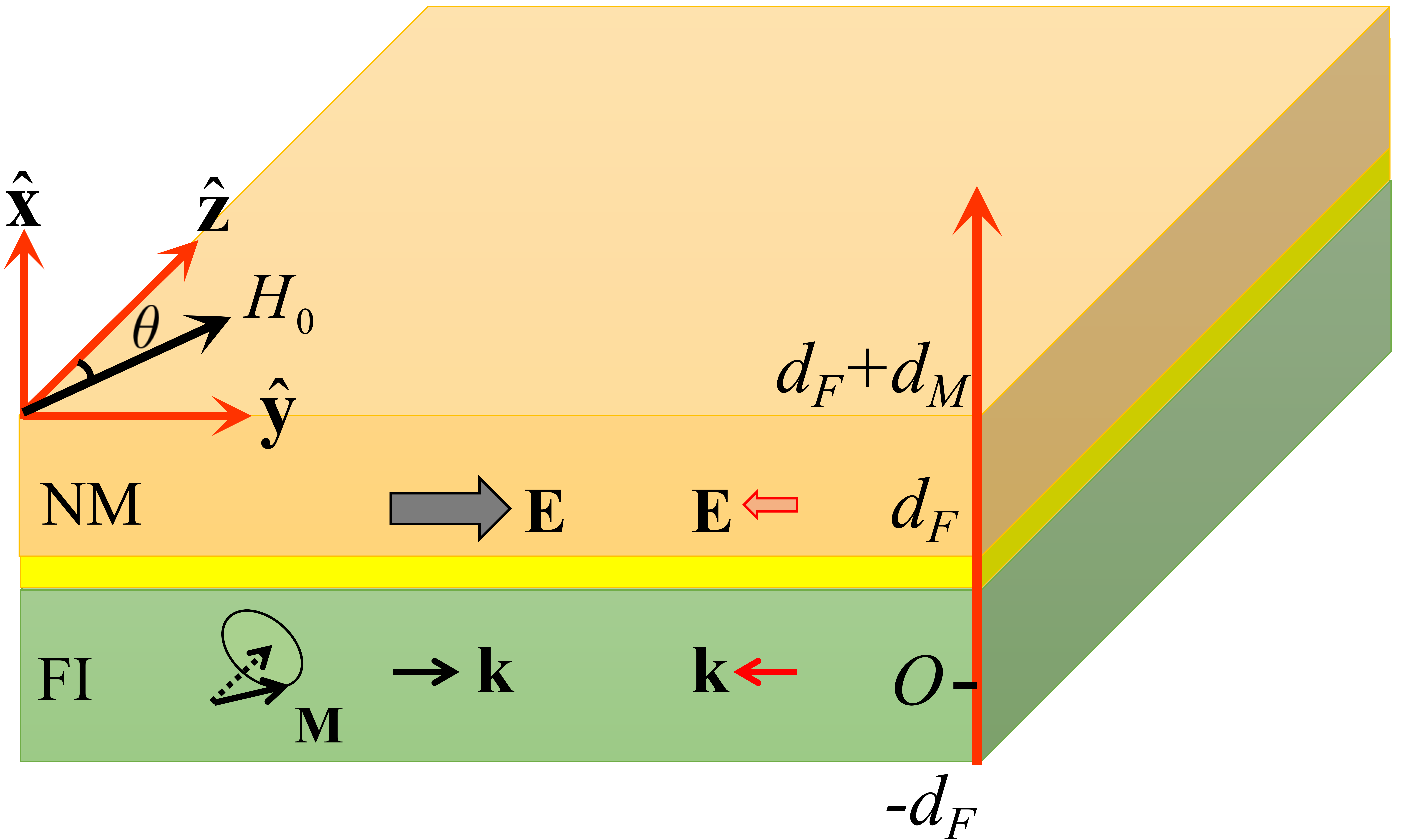}
\caption{Ferromagnetic insulator-normal metal heterostructure. An in-plane external magnetic field ${\bf H}_0$ orients the magnetization at an angle $\theta$ with the $\hat{\bf z}$-direction. The yellow sheet between the normal metal and ferromagnetic insulator indicates suppression of the exchange interaction and conventional spin pumping.}
\label{model}
\end{figure}

In this Letter,  we address the counter-intuitive effect that the strong momentum-dependent eddy-current-induced damping by a normal metal overlayer as shown in Fig.~\ref{model} may help surmount obstacles such as magnetic inhomogeneities~\cite{surface_roughness}, artificial potential barriers formed by surface scratches~\cite{scratches}, or dc-current carrying wires~\cite{semitransparent}. Here we focus on the band edges of magnetic films that are much thinner than the extinction length of the Damon-Eshbach surface states in thick slabs and are therefore not chiral. Instead, the effect therefore originates from the Oersted fields generated by the eddy currents in the overlayer that act in only half of the reciprocal space~\cite{Yu_chirality} and causes magnon accumulations at the sample edges or magnon skin effect~\cite{non_Hermitian_magnons,Flebus}. The transmission through a barrier that is small and symmetric for magnons with opposite wave numbers in an uncovered sample becomes unidirectional with the assistance of dissipative eddy currents.

\textit{Chiral-damping-enhanced wave transmission}.---As a simple illustration of the general physics, we start with considering a 1D wave field with a single short-range scattering potential $V \delta \left(  x\right)$ at the origin. The  dimensionless wave or Schr\"{o}dinger equation for the scattering states with kinetic energy $E=k^{2}>0$ reads $\left(  -\partial^{2}+V\delta\left(  x\right)  -k^{2}\right)  \psi\left(
x\right)  =0$.
Next, we assume a chiral damping that affects only the right-moving waves. We use a Gilbert type of viscous damping proportional to the velocity with the parameter $\eta$.  The solutions to this problem are linear combinations
of plane waves. States \textit{coming from the right} then read
\begin{equation}
\psi_{R}^{\mathrm{(c)}}=\left\{
\begin{array}
[c]{c}%
e^{-ikx}+r^{\mathrm{(c)}}e^{ikx}e^{-\eta kx}\\
t^{\mathrm{(c)}}e^{-ikx}%
\end{array}
\right.  \text{\, for  \,\,}%
\begin{array}
[c]{c}%
x>0\\
x<0
\end{array},
\end{equation}
where $r/t$ are the reflection/transmission coefficients that solve the scattering problem. The transmission and reflection probabilities read
\begin{subequations}
\begin{align}
\left\vert t^{\mathrm{(c)}}\right\vert ^{2}  & =\frac{\left(  \eta k\right)  ^{2}+4k^{2}%
}{\left(  V+\eta k\right)  ^{2}+4k^{2}},\\
\left\vert r^{\mathrm{(c)}}\right\vert ^{2}&=\frac{V^{2}}{\left(  V+\eta k\right)  ^{2}
+4k^{2}}.
\end{align}
\end{subequations}
When the damping is sufficiently large $(\eta k\gg\left\vert V\right\vert )$, the
transmission probability becomes unity and reflection vanishes,
irrespective of the scattering potential, i.e., which in essence is the anomalous transmission reported below for a magnetic device.  On the other hand, for non-chiral damping, e.g., $\psi_{L}^{\left(  0\right)  } =e^{ikx}e^{-\eta kx}$ and $\psi_{R}^{\left(  0\right)  } =e^{-ikx}e^{\eta kx}$, the states
coming from the right read
\begin{equation}
\psi_{R}^{\mathrm{(nc)}}=\left\{
\begin{array}
[c]{c}%
\left(  e^{-ikx}+r^{\mathrm{(nc)}}e^{ikx}\right)  e^{-\eta kx}\\
t^{\mathrm{(nc)}}e^{-ikx}e^{\eta kx}%
\end{array}
\right.  \text{\, for  \,\,        }%
\begin{array}
[c]{c}%
x>0\\
x<0
\end{array},
\end{equation}
for which the transmission and reflection probabilities
\begin{subequations}
\begin{align}
|t^{\mathrm{(nc)}}|^{2}=\frac{4k^{2}}{(2\eta k+V)^{2}+4k^{2}}\overset{\eta\gg1}{\rightarrow}0,\\
|r^{\mathrm{(nc)}}|^{2}=\frac{\left(  V+2\eta k\right)  ^{2}}{(2\eta k+V)^{2}+4k^{2}%
}\overset{\eta\gg1}{\rightarrow}1.
\end{align}
\end{subequations}
Therefore, conventional non-chiral damping \textit{suppresses} the transmission. 
The enhanced transmission is caused by reducing the energy cost of an asymmetric wave function curvature at the scatterer. The principle should hold for wave propagation in arbitrary systems with non-reciprocal damping. In the following, we focus on a magnetic device in which the effect should be observable.

\textit{Model}.---To be specific, we consider the ferromagnetic insulator (FI)-normal metal (NM) heterostructure with thickness $2d_F$ and $d_M$ and an in-plane magnetic field ${\bf H}_0$ in Fig.~\ref{model}.  The saturated equilibrium magnetization ${\bf M}_s$ makes an angle $\theta$ with the $\hat{\bf z}$-direction such that the torques exerted by the external and anisotropy fields cancel. For convenience, we set $\theta=0$ in the following discussion and defer results for finite $\theta$  to the Supplemental Material (SM)~\cite{supplement}. We generalize a previous adiabatic theory~\cite{eddy_currents1,Yu_chirality} to the full electrodynamics of the system by self-consistently solving the Maxwell equations coupled with the linearized Landau-Lifshitz (LL) equations and Ohm's Law. This treatment becomes exact in the limit of an instantaneous response of the metal electrons and high-quality ultrathin magnetic films.

The driving force is an externally generated spatiotemporal magnetization dynamics \( \mathbf{M} (\mathbf{r},t) =\mathbf{M} (\mathbf{r}, \omega)e^{-i \omega t}\)  at frequency $\omega$.  According to Maxwell's theory, the electric field \textbf{E}  obeys the wave equation
$\nabla^2 \mathbf{E}(\mathbf{r}, \omega)+k_0^2 \mathbf{E}(\mathbf{r}, \omega)=-i\omega \mu_0  {\bf J}_M$, where the wave number $k_0=\omega\sqrt{\mu_0\varepsilon_0}$, $\mu_0$  $(\varepsilon_0)$ is the vacuum permeability (permittivity), and  ${\bf J}_M=\nabla\times{\bf M}$ is the ``magnetization current"~\cite{Jackson}. Disregarding the intrinsic Gilbert damping, the LL equation
$i\omega \mathbf{M}=-\mu_0 \gamma \mathbf{M} \times \mathbf{H}_{\rm eff}[\textbf{M}]$
governs the magnetization dynamics in the FI, where $\gamma$ is the gyromagnetic ratio. The effective magnetic field $\mathbf{H}_{\rm eff}[\mathbf{M}]=-\delta F [\mathbf{M}]/ \delta \mathbf{M}(\mathbf{r})$, where the free energy $F$ is a functional of the magnetization. It includes the static field ${\bf H}_0$, the dipolar field ${\bf H}_d$,  and (in the FI) the exchange field ${\bf H}_{\rm ex}=\alpha_{\rm ex}\nabla^2 {\bf M}$ that depends on the spin-wave stiffness $\alpha_{\rm ex}$. In the presence of the NM layer, $\mathbf{H}_{\rm eff}[\textbf{M}]$ also contains the Oersted magnetic fields generated by the ``eddy" currents \(\mathbf J=\sigma \mathbf E\), where the electrical conductivity \(\sigma\) is real. This defines a closed self-consistency problem that we solve numerically.

We consider a thin FI film with constant ${\bf M}_s=(0,0,M_s)$. The transverse fluctuations 
${\bf M}(\mathbf{r}, \omega)=(M_x(\mathbf{k}, \omega),M_y(\mathbf{k}, \omega),0)e^{i \mathbf{k} \cdot \mathbf{r}}$ with in-plane wave vectors $\mathbf{k}=(0, k_y, k_z)$ are small precessions with $i M_x(\mathbf{k}, \omega)=a_{\bf k} M_y(\mathbf{k},\omega)$, where the complex ellipticity $a_{\bf k}$ becomes unity for circular motion.

The electric-field modes outside the magnet are plane waves with 
wave numbers $k_{m}=\sqrt{\omega^2\mu_0\varepsilon_0 +i\omega \mu_0  \sigma}$, where $\sigma = 0$ in the absence of an NM layer. The continuity of electric and magnetic fields provides the interface boundary conditions. The field in the FI
\begin{align}
&E_{\eta=\{x,y,z\}}({-d_F \leqslant x \leqslant d_F})\nonumber\\
&=E^{(0)}_{\eta}({-d_F \leqslant x \leqslant d_F})+{\cal R}_kE^{(0)}_{\eta}({x=d_F}) e^{-i A_k\left(x-d_F\right)}
\nonumber
\end{align}
is now modified by the reflection coefficient
\begin{align}
{\cal R}_k=\frac{\left(A_k^2-B_k^2\right) e^{i B_k d_M}-\left(A_k^2-B_k^2\right) e^{-i B_k d_M}}{(A_k-B_k)^2 e^{i B_k d_M}-(A_k+B_k)^2 e^{-i B_k d_M}}, 
\end{align}
where $E^{(0)}$ is the solution inside the FI without the NM cap~\cite{supplement}, $A_k=\sqrt{k_{0}^2-k^2}$, and $B_k=\sqrt{k_{m}^2-k^2}$. The reflection is isotropic and strongly depends on the wave vector. Naturally, ${\cal R}_k=0$ when $d_M=0$. 
On the other hand, when $|{\bf k}|=0$,  the electric field cannot escape the FI, since the reflection is total with ${\cal R}_k=-1$.

A corollary of Maxwell's equation---Faraday's Law--- reads in frequency space $i\omega \mu_0[{\bf H}_d(\mathbf{r},\omega)+{\bf M}(\mathbf{r},\omega)]=\nabla \times {\bf E}(\mathbf{r},\omega)$.
When the magnetization of sufficiently thin magnetic films is uniform, the Zeeman interaction is proportional to the spatial average ${\bf H}_d$ over the film thickness. Referring to the SM for details~\cite{supplement}, we find
\begin{align}
{H}_{d,x}&=\left[-\frac{{\cal R}_k}{4A_k^2d_Fa_{\bf k}}(e^{2iA_kd_F}-1)^2(-iA_ka_{\bf k}+k_y)\right.\nonumber\\
&\left.+\frac{i}{2A_kd_F}(e^{2iA_kd_F}-1)\right]M_x\equiv\zeta_x({\bf k})M_x,\nonumber
\end{align}
\begin{align}
{H}_{d,y}&=\left[-\frac{{\cal R}_k}{4iA_kd_F}(e^{2iA_kd_F}-1)^2\left(\frac{-k_y}{iA_k}a_{\bf k}+\frac{k_y^2}{A_k^2}+2\right)\right.\nonumber\\
&\left.+\frac{k_y^2}{A_k^2}-\frac{k_y^2}{A_k^2}\frac{1}{2iA_kd_F}(e^{2iA_kd_F}-1)\right]M_y\equiv \zeta_y({\bf k})M_y.
\nonumber
\end{align}
By substitution into the LL equation, the spin wave eigenfrequencies and ellipticities become 
\begin{subequations}
\begin{align}
\omega({\bf k})&=\mu_0 \gamma \sqrt{(\tilde{H}_0-\zeta_x({\bf k})M_s)(\tilde{H}_0-\zeta_y({\bf k})M_s)},\label{dispersion}\\
a_{\bf k}&=\sqrt{(\tilde{H}_0-\zeta_y({\bf k})M_s)/(\tilde{H}_0-\zeta_x({\bf k})M_s)},
\end{align}
\end{subequations}
where $\tilde{H}_0=H_0+\alpha_{\rm ex}k^2M_s$. $\mathrm{Im}\, \omega({\bf k})\ne 0$ because of the Joule heating due to the eddy currents in the cap layer.

\begin{figure}[htbp]
\includegraphics[width=0.48\textwidth,trim=0.3cm 0cm 0cm 0.1cm]{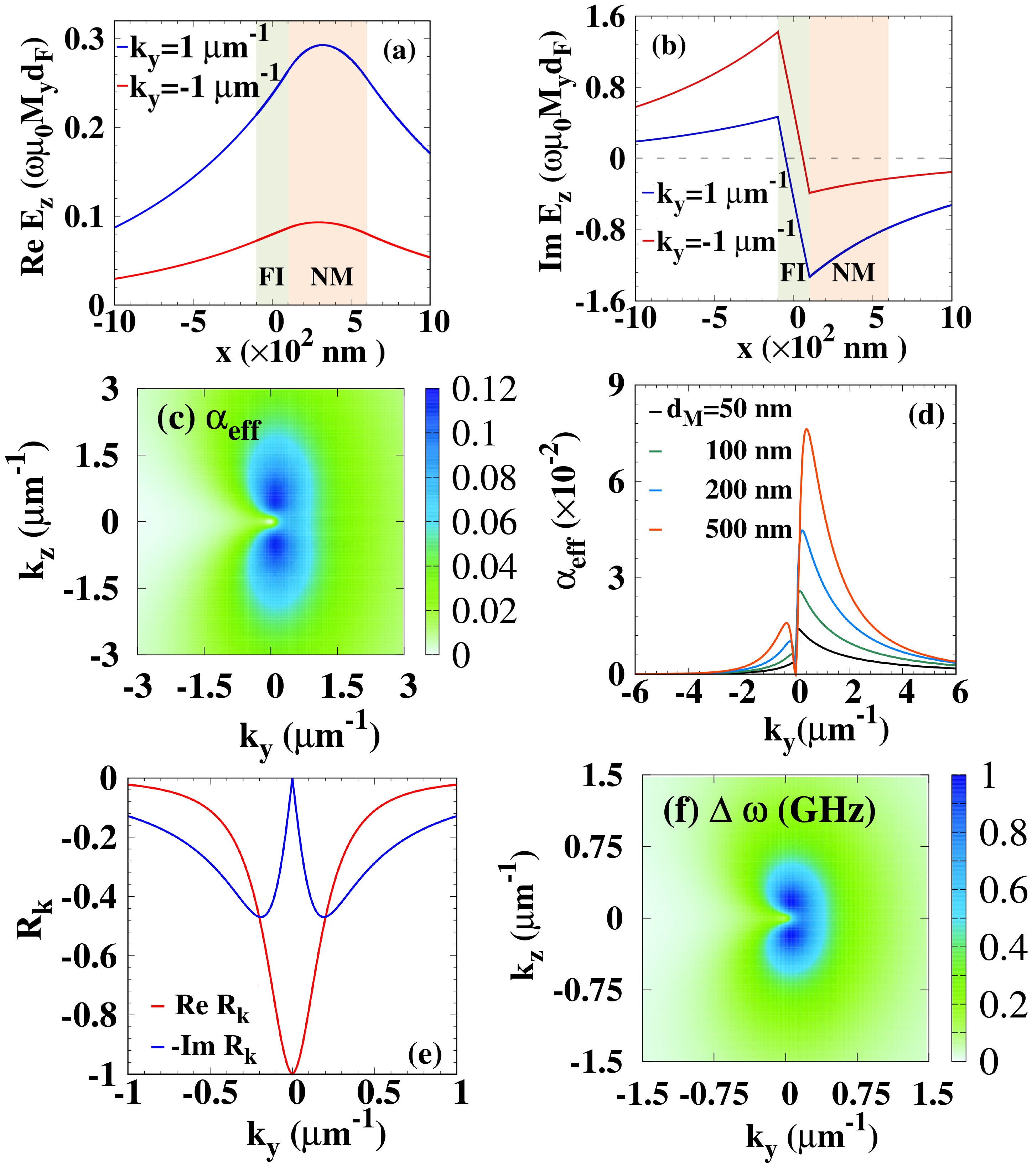}
\caption{The system responds strongly to a phase difference between the spin waves and their wave vector-dependent ac electric stray fields \(\textbf{E}\). Re\(\textbf{E}\) causes damping [(a)] and Im\(\textbf{E}\)  a frequency shift [(b)]. 
\(\mathrm{Re} \, E_z\) governs the spin wave vector dependence of the chiral damping [(c)]. (d) illustrates the strong $k_y$-dependence of the damping of the lowest standing spin wave for Cu thicknesses $d_M=\{50, 100, 200, 500\}$~nm. (e) shows the real and imaginary parts of the reflection coefficient ${\cal R}_k$ that causes the frequency shifts plotted in (f).}
\label{damp}
\end{figure}

\textit{Chiral damping and frequency shifts}.---The stray electric fields of spin waves propagating perpendicular to the magnetization are chiral, \textit{i.e.}, they depend on their propagation direction by a hand rule. When $k_z=0$, ${\bf E}=E_z\hat{\bf z}$ is along the equilibrium magnetization and $E_z\propto M_y$ is complex and chiral, i.e. non-vanishing for positive $k_y$.  We illustrate the results of the self-consistent calculations for  $d_F=100$~nm, $d_M=500$~nm,  conductivity $\sigma=6.0 \times 10^{7}~(\rm{\Omega\cdot m})^{-1}$ for copper at room temperature \cite{Griffiths}, applied magnetic field $\mu_0H_0=0.02$~T, $\mu_0 M_s=0.178$~T, the exchange stiffness
$\alpha_{\rm ex}=3 \times 10^{-16}~\rm{m}^2$ for YIG \cite{YIG1}, and $\gamma=1.77 \times 10^{11} ~(\rm{s\cdot T})^{-1}$. The presence of the NM cap layers shifts the relative phases between the stray electric fields and that of the generating spin waves. We focus here on the wave numbers $k_y=\pm 1~\rm{\mu m}^{-1}$ in Fig.~\ref{damp}(a) [Fig.~\ref{damp}(b)] at which the electric field is in-phase (out-of-phase) with the transverse magnetization $M_y\hat{\bf y}$.  The response to an in-phase (out-of-phase) electric field is dissipative (reactive). Both components decay in the FI and the vacuum as $\propto 1/|\bf k|$. In the NM, the in-phase component is screened only in the metal region on the scale of a skin depth $\lambda= \sqrt{2/(\omega \mu_0 \sigma)} \sim 1.5~ \rm{\mu}$m at $\omega=11$ GHz. The out-of-phase electric field, on the other hand, creates only a reactive response.
Also in this case the damping is modulated for constant Gilbert damping by the associated spin wave frequency shift in Fig.~\ref{damp}(b), an effect that cannot be captured by the adiabatic approximation~\cite{eddy_currents1,Yu_chirality}.

The chirality of the radiated electric field controls the backaction of the NM layer that modifies the magnon dispersion in a chiral fashion. Figure~\ref{damp}(c) illustrates the strong wave vector-dependent  damping coefficient $\alpha_{\rm eff}({\bf k})=|{\rm Im} \, {\omega}_{\bf k}|/{\rm Re}\,\omega_{\bf k}$. Spin waves propagating in the positive $\hat{\bf y}$-direction decay much faster than those along the negative direction, while the damping for positive and negative $k_z$ is the same. 
According to Fig.~\ref{damp}(d), the calculated damping for $k_z=0$ in Fig.~\ref{damp}(c) increases with the thickness of the Cu (YIG) film~\cite{supplement}. The enhancement of the damping saturates for NM thicknesses  $d_N > 1/\sqrt{k^2+1/\lambda^2}$, depending on the skin depth ($\lambda \sim 1.5 ~\rm{\mu m}$)  and the wave number $1/k$ of the electric field. Moreover, the Kittel mode at ${\bf k}=0$  in Fig.~\ref{damp}(e) is not affected by the metal at all because the reflection coefficient  ${\cal R}_k=-1$, which implies that the dynamics of the FI and metal fully decouple.  Indeed, recent experiments do not find a frequency shift of the FMR by a superconducting overlayer~\cite{FS_no_shift_1,FS_no_shift_2}. The additional damping by eddy currents reported by Ref.~\cite{eddy_currents4} is caused by the width of the exciting coplanar waveguide, a finite-size effect that we do not address here.

The real part of ${\cal R}_k$ in Fig.~\ref{damp}(e) causes an in-phase Oersted magnetic field that chirally shifts the spin wave frequencies by as much as $\sim 1$~GHz, see Fig.~\ref{damp}(f). Reference~\cite{Beach} indeed reports a frequency shift of perpendicular standing spin wave modes in Bi-YIG films in the presence of thin metallic overlayers.

The predicted effects differ strongly from those caused by spin pumping due to the interface exchange coupling  $\alpha_{\rm sp}=(\hbar \gamma/M_s d_F) \mathrm{Re} \,g_{\uparrow\downarrow}$,  where $g_{\uparrow\downarrow}$ is the interfacial spin mixing conductance~\cite{spin_pumping}. $\alpha_{\rm sp}$ does not depend on the thickness of the metal and vanishes like \(1/d_F\). The frequency shift scales like \(\mathrm{Im} \,g_{\uparrow\downarrow}/d_F\) and is very small even for ultrathin magnetic layers. In contrast, the eddy current-induced damping is non-monotonic, scaling like \(\propto d_F\) when $2kd_F \ll 1$, vanishing for much thicker magnetic layers, and reaching a maximum at $d_F \sim 2 \lambda$.

\textit{Unidirectional transmission of wave packets through a potential barrier}.---The transmission of a wave packet impinging from the left or right at a conventional potential barrier is the same~\cite{quantum_transport_Yuli}. In the presence of a metal cap, this does not hold for magnons in thin magnetic films. 

A quantum mechanical formalism is en vogue when describing non-Hermitian systems since it is convenient to keep track of phases. The equations hold for classical systems when replacing the operators by field amplitudes. We illustrate the effect of a square potential barrier of width $d$ and height $u_0$, $\hat{V}(y)=u_0[\Theta(y+d/2)-\Theta(y-d/2)]$ where $\Theta(x)$ is the Heaviside step function, on the magnon transmission along  $\hat{\bf y}(\perp {\bf M}_s)$. 
With incoming $\langle y|k_0\rangle=e^{ik_0y}$, the scattered states $|\psi_s\rangle$ obey the Lippmann-Schwinger formula~\cite{Sakurai} 
\begin{align}
    |\psi_s\rangle=|k_0\rangle+\frac{1}{i\hbar \partial_t-\hat{H}_0+i0_+}\hat{V}|\psi_s\rangle.
\end{align}
where $\hat{H}_0=\sum_{\bf k}\hbar \omega_{\bf k} \hat{m}_{\bf k}^{\dagger}\hat{m}_{\bf k}$ is the magnon Hamiltonian for an extended film and $\hat{m}_{\bf k}$ is the annihilation operator of magnons with frequency $\omega_{\bf k}$  from Eq.~\eqref{dispersion}.
The transmitted waves read
\begin{align}
    \langle y|\psi_s\rangle=\left\{\begin{array}{cc}
    T_{+}(k_0)e^{ik_0y},~~~~~\{y,k_0\}>0\\
    T_{-}(k_0)e^{ik_0y},~~~~~\{y,k_0\}<0
    \end{array}\right..
\end{align}
In the weak scattering limit $|u_0d|\ll |\hbar v_{k_0}|$, 
\begin{align}
    T_{\pm}(k_0)&=1 \pm \left(\frac{i\hbar v_{k_0}}{u_0 d}-\frac{v_{k_0}}{2|v_{k_0}|}\right)^{-1}\approx 1\mp i\frac{u_0d}{\hbar v_{k_0}},
\end{align}
where $v_{k_0}=\partial \omega_{\bf k}/\partial {\bf k}|_{{\bf k}=k_0\hat{\bf y}}$ is a generalized group velocity that dissipation renders complex. The imaginary part of the group velocity and transmission amplitudes depend on the direction of the incoming wave:
\begin{align}
    D_{\pm}(k_0)&=|T_{\pm}(k_0)|^2\approx 1 \pm 2{\rm Im} \left(\frac{u_0d}{\hbar v_{k_0}}\right).
    \label{transmission_analytic}
\end{align}
For example, with $u_0/\hbar=30.5$~GHz, $d=0.1~\mu {\rm m}$, $k_0=\pm 0.8~\mu {\rm m}^{-1}$, $v_{k_0>0}=(2.32+0.52i)$~km/s and $v_{k_0<0}=-(2.64+0.16i)$~km/s lead to ${\cal T}_+(k_0>0)\approx 0.6$ while  ${\cal T}_-(k_0<0)\approx 0.9$, so even in the weak scattering limit the NM cap layer significantly and asymmetrically reduces the transmission probability.

We then numerically address the potential scattering for a two-dimensional square lattice with $\hat{m}_i=(1/\sqrt{N})\sum_{\bf k} \hat{m}_{\bf k} e^{i{\bf k}\cdot{\bf r}_i}$, where $i$ labels the sites and $N$ is the number of sites. The Hamiltonian in the real space
$\hat{H}_0=\sum_{ij} t_{ji} \hat{m}_j^{\dagger}\hat{m}_i$,
where $t_{ji}=({1}/{N})\sum_{\bf k}\hbar 
\omega_{\bf k}e^{i{\bf k}\cdot({\bf r}_j-{\bf r}_i)}$ is a hopping amplitude between possibly distant sites \textit{i} and \textit{j} and the summation is over the first Brillouin zone. With a coarse-grained lattice constant of $a_y=a_z=0.1~\mu$m the reciprocal lattice vector $2\pi/a_{y,z}$ is much larger than the wave numbers of the magnon modes of interest (refer to the SM~\cite{supplement} for details). When the frequencies   $\omega_{\bf k}$  are complex, the Hamiltonian is non-Hermitian, \textit{i.e.},  
$t_{ji}\ne t_{ij}^*$. As derived in the SM~\cite{supplement}, magnons propagate in the negative direction without damping but decay when propagating in the opposite one. Their accumulation at the left boundary of the sample is a non-Hermitian skin effect~\cite{skin1,skin2,skin3,Bergholtz}.

\begin{figure}[htb]
\includegraphics[width=0.47\textwidth,trim=0cm 0cm 0cm 0.1cm]{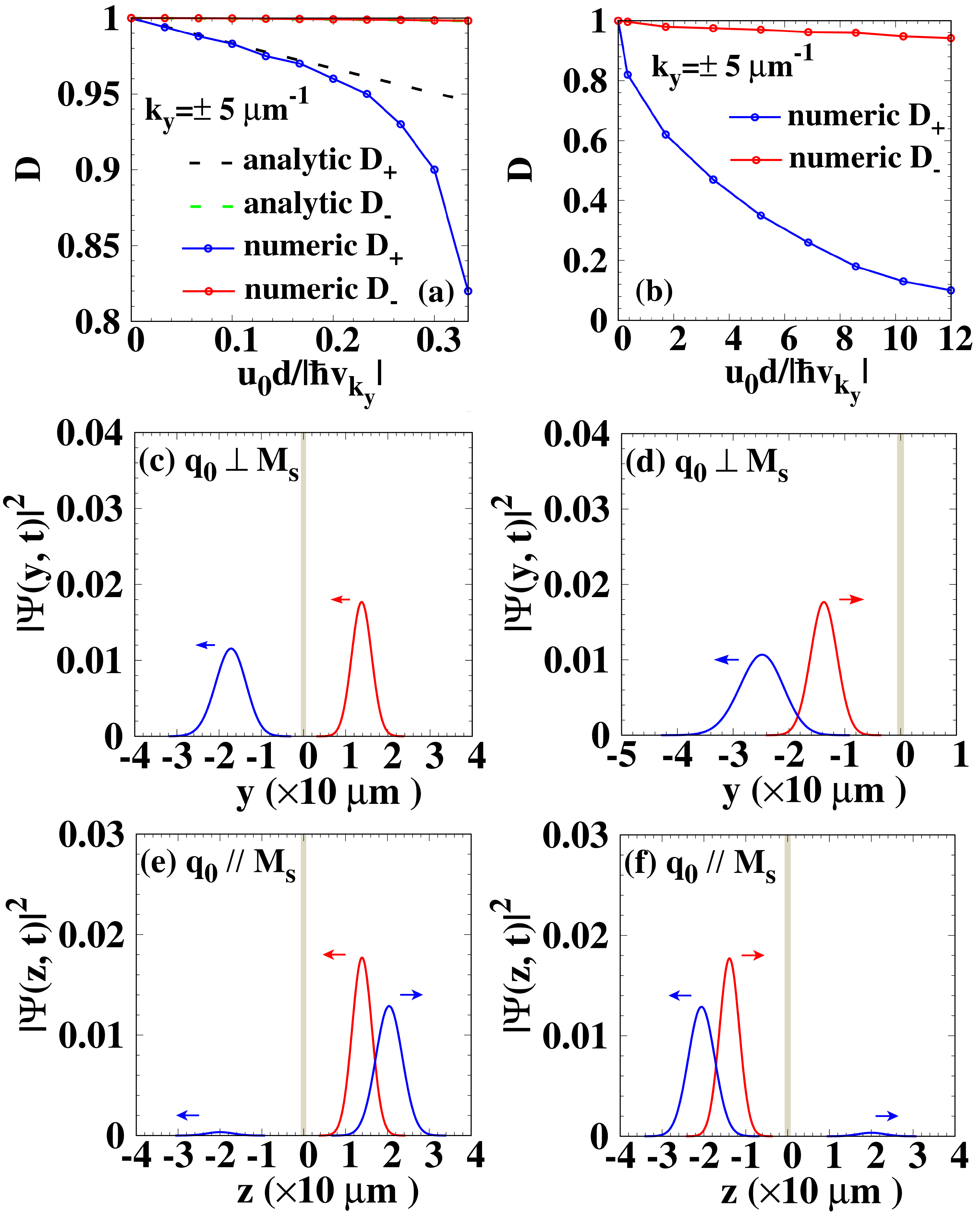}
        \caption{Calculated transmissions [(a) and (b)] and time evolution of spin-wave packets in the presence of a potential barrier at the origin when ${\bf q}_0\perp {\bf M}_s$ [(c) and (d)] and ${\bf q}_0\parallel {\bf M}_s$ [(e) and (f)], where \({\bf M}_s\) and the applied magnetic field are parallel to the sample edge with $\theta=0$.  The red and blue curves represent, respectively, the incident and scattered wave packets with propagation directions indicated by arrows. }
        \label{scatter}
\end{figure}

We can now assess the strong scattering regime with $|u_0d |\gtrsim |\hbar v_{k_0}|$ by numerical calculations to find dramatic effects on the time evolution of a real-space spin-wave packet as launched,  \textit{e.g.}, by a current pulse in a microwave stripline. We adopt a Gaussian shape 
$\Psi({\bf r},0)=e^{-({\bf r}-{\bf r}_0)^2/(2\eta^2)}e^{i{\bf q}_0\cdot{\bf r}}$ centered at ${\bf r}_0$ with a width $\eta\gg a_{y,z}$ that envelopes a plane wave with wave vector ${\bf q}_0$ and 
$\hat{V}({\bf r})=u_0 f({\bf r})$ with either   $f(|y-\tilde{y}_0|<d)=1$ or $f(|z-\tilde{z}_0|<d)=1$, where $\tilde{y}_0$ and $\tilde{z}_0$ are the center of the barriers, and zero otherwise. According to Schr$\ddot{\rm o}$dinger's equation 
$\Psi({\bf r},t)=e^{i\hat{H}t/\hbar}\Psi({\bf r},t=0)$ with  $\hat{H}=\hat{H}_0+\hat{V}({\bf r})$. Numerical results in Fig.~\ref{scatter}(a) and (b)   $u_0d\ll |\hbar v_{k_0}|$  agree with perturbation theory (\ref{transmission_analytic}) in the weak scattering regime. However, when $|\hbar v_{k_0}|\lesssim u_0d$ and $|{\rm Im}(v_{-k_0})|\ll|{\rm Im}(v_{k_0})|\lesssim |{\rm Re}(v_{\pm k_0})|$ the transmission and unidirectionality becomes almost perfect. 
Figure~\ref{scatter}(c) and (d) show a nearly unidirectional transmission of the wave packet through the potential barrier for the Damon-Eshbach configuration ${\bf q}_0\perp {\bf M}_s$; it is transparent for spin waves impinging from the left, but opaque for those from the right.  In the calculations, ${\bf q}_0=q^{(0)}_{y}\hat{\bf y}$ with $q^{(0)}_{y}=\pm 5~\mu {\rm m}^{-1}$ and $\eta=3~\mu {\rm m}\gg d$. The potential barrier is peaked with $d=a_{y,z}=0.1~\mu {\rm m}$ and its height $u_0/\hbar=15$~GHz is relatively weak (the regular on-site energy $\sim 13$~GHz). Also, $d_M=100$~nm and $d_F=100$~nm. The results are insensitive to the detailed parameter values (see the SM~\cite{supplement}). 
The red and blue curves are the incident and reflected wave packets, respectively. When $q^{(0)}_y<0$, the barrier does not affect the wave packet that propagates freely through the potential barrier and accumulates on the left edge  [Fig.~\ref{scatter}(c)]. When $q^{(0)}_y>0$, as shown in Fig.~\ref{scatter}(d), the barrier reflects the wave packet nearly completely, demonstrating the chiral-damping-enhanced wave transmission.  

Since the chiral damping is crucial, its absence in waves propagating in the $\hat{\bf z}$-direction must affect the transport over the barrier.  Indeed,  our calculations in Fig.~\ref{scatter}(e) and (f) find strong reflection for both propagation directions, even when reducing the barrier height by an order of magnitude to $u_0=1.5$~GHz (see the SM~\cite{supplement}). 

\textit{Discussion and conclusion}.---We calculate the chiral damping, chiral frequency shift, and anomalous transport of magnonic modes in ferromagnetic films with NM cap layers beyond the adiabatic approximations. We predict anomalous unidirectional spin transport over potential barriers. Thereby, chiral damping negates the conventional wisdom that in the linear transport regime the transmission of waves (electrons, phonons, photons, and magnons) through an obstacle is symmetric, i.e., independent of the direction of incidence. We illustrate the general principle for a realistic system of practical importance. In this sense, the damping does not cause harm but has beneficial effects that improve functionality and enhance mobility. Our predictions are not limited to magnons, but carry over for the propagation of all chiral quasiparticles, such as surface acoustic waves~\cite{SAWs_1,SAWs_2}, microwaves in loaded waveguides with magnetic insertions~\cite{Canming,Lin}, or chiral waveguides for light~\cite{chiral_photonics_1,chiral_photonics_2}.

\begin{acknowledgments}
This work is financially supported by the National Key Research and Development Program of China under Grant No.~2023YFA1406600, the National Natural Science Foundation of China under Grants No.~12374109 and No.~12088101, the startup grant of Huazhong University of Science and Technology, as well as JSPS KAKENHI Grants No. 19H00645, 22H04965, and JP24H02231.
\end{acknowledgments}


\begin{thebibliography}{99}

\bibitem{Lenk} B. Lenk, H. Ulrichs, F. Garbs, and M. M\"{u}nzenberg, The building blocks of magnonics, Phys. Rep. \textbf{507}, 107 (2011).

\bibitem{Chumak} A. V. Chumak, V. I. Vasyuchka, A. A. Serga, and B. Hillebrands, Magnon spintronics, Nat. Phys. \textbf{11}, 453 (2015).

\bibitem{Grundler} D. Grundler, Nanomagnonics around the corner, Nat. Nanotechnol. \textbf{11}, 407 (2016).

\bibitem{Demidov} V. E. Demidov, S. Urazhdin, G. de Loubens, O. Klein, V. Cros, A. Anane, and S. O. Demokritov, Magnetization oscillations and waves driven by pure spin currents, Phys. Rep. \textbf{673}, 1 (2017).


\bibitem{Brataas} A. Brataas, B. van Wees, O. Klein, G. de Loubens, and M. Viret, Spin Insulatronics, Phys. Rep. \textbf{885}, 1 (2020).

\bibitem{Barman} A. Barman, G. Gubbiotti, S. Ladak, A. O. Adeyeye, M. Krawczyk, J. Gr{\"a}fe, C. Adelmann, S. Cotofana, A. Naeemi,
V. I. Vasyuchka \textit{et al}., The 2021 magnonics roadmap, J. Phys.:
Condens. Matter \textbf{33}, 413001 (2021).


\bibitem{Yu_chirality} T. Yu, Z. C. Luo, and G. E. W. Bauer, Chirality as generalized spin–orbit interaction in spintronics, Phys. Rep. \textbf{1009}, 1 (2023).

\bibitem{non_Hermitian_magnons} T. Yu, J. Zou, B. Zeng, J. W. Rao, and K. Xia, Non-Hermitian topological magnonics, Phys. Rep. \textbf{1062}, 1 (2024).

\bibitem{Flebus} H. M. Hurst and B. Flebus, Non-Hermitian physics in magnetic systems, J. Appl. Phys. \textbf{132}, 220902 (2022).

\bibitem{magnon_memory} K. Baumgaertl and D. Grundler, Reversal of nanomagnets by propagating magnons in ferrimagnetic yttrium iron garnet enabling nonvolatile magnon memory, Nat. Commun. \textbf{14}, 1490 (2023).








\bibitem{study_1} M. Eschrig, Spin-polarized supercurrents for spintronics: a review of current progress, Rep. Prog. Phys. \textbf{78}, 104501 (2015).

\bibitem{study_2} J. Linder and J. W. A. Robinson, Superconducting spintronics, Nat. Phys. \textbf{11}, 307 (2015).

\bibitem{Bergeret} F. S. Bergeret, M. Silaev, P. Virtanen, and T. T. Heikkil\"a, Colloquium: Nonequilibrium effects in superconductors with a spin-splitting field,
Rev. Mod. Phys. \textbf{90}, 041001 (2018).


\bibitem{Weihan} R. Cai, I. \ifmmode \check{Z}\else \v{Z}\fi{}uti\ifmmode \acute{c}\else \'{c}\fi{}, and W. Han, Superconductor/ferromagnet heterostructures: a platform for superconducting spintronics and quantum computation, Adv. Quan. Technol. \textbf{6}, 2200080 (2023).
\bibitem{Q_information} Y. Tabuchi, S. Ishino, A. Noguchi, T. Ishikawa,   R. Yamazaki,  K. Usami, and  Y. Nakamura, Coherent coupling between a ferromagnetic magnon and a superconducting qubit, Science \textbf{349}, 405 (2015).

\bibitem{Q_information_1} D. L. Quirion, S. P. Wolski, Y. Tabuchi, S. Kono, K. Usami, and Y. Nakamura, Entanglement-based single-shot detection of a single magnon with a superconducting qubit, Science \textbf{ 367}, 425 (2020).


\bibitem{Q_information_2} T. Yu, M. Claassen, D. M. Kennes, and M. A. Sentef, Optical manipulation of domains in chiral topological superconductors, Phys. Rev. Research \textbf{3}, 013253 (2021).

\bibitem{Q_information_3}
H. Y. Yuan, Y. Cao, A. Kamra, R. A. Duine, and P. Yan, Quantum magnonics: When magnon spintronics meets quantum information science, Phys. Rep. \textbf{965}, 1 (2022).


\bibitem{Q_information_6}
Z. Li, M. Ma, Z. Chen, K. Xie, and F. Ma, Interaction between magnon and skyrmion: Toward quantum magnonics, J. Appl. Phys. \textbf{132}, 210702 (2022).


\bibitem{Q_information_4} B. Z. Rameshti, S. V. Kusminskiy, J. A. Haigh, K. Usami, D. Lachance-Quirion, Y. Nakamura, C. -M. Hu, H. X. Tang, G. E. W. Bauer, and Y. M. Blanter, Cavity magnonics, Phys. Rep. \textbf{979}, 1 (2022).

\bibitem{Q_information_5} D. Xu, X.-K. Gu, H.-K. Li, Y.-C. Weng, Y.-P. Wang, J. Li, H. Wang, S.-Y. Zhu, and J. Q. You, Quantum Control of a Single Magnon in a Macroscopic Spin System, Phys. Rev. Lett. \textbf{130}, 193603 (2023).


\bibitem{topological} K. M. D. Hals, M. Schecter, and M. S. Rudner, Composite Topological Excitations in Ferromagnet-Superconductor Heterostructures,
Phys. Rev. Lett. \textbf{117}, 017001 (2016).


\bibitem{topological_6}
M. Kargarian, D. K. Efimkin, and V. Galitski, Amperean Pairing at the Surface of Topological Insulators,
Phys. Rev. Lett. \textbf{117}, 076806 (2016).

\bibitem{topological_4}
N. Rohling, E. L. Fj{\ae}rbu, and A. Brataas, Superconductivity induced by interfacial coupling to magnons,
Phys. Rev. B \textbf{97}, 115401 (2018).
\bibitem{topological_5}
H. G. Hugdal, S. Rex, F. S. Nogueira, and A. Sudb{\o}, Magnon-induced superconductivity in a topological insulator coupled to ferromagnetic and antiferromagnetic insulators, Phys. Rev. B \textbf{97}, 195438 (2018).

\bibitem{topological_3}
E. L. Fj{\ae}rbu, N. Rohling, and A. Brataas, Superconductivity at metal-antiferromagnetic insulator interfaces, Phys. Rev. B \textbf{100}, 125432 (2019).


\bibitem{topological_2} K. M\ae{}land  and A. Sudb\o{}, Topological Superconductivity Mediated by Skyrmionic Magnons, 
Phys. Rev. Lett. \textbf{130}, 156002 (2023).

\bibitem{superconductor_gating_theory} T. Yu and G. E. W. Bauer, Efficient Gating of Magnons by Proximity Superconductors, Phys. Rev. Lett. \textbf{129}, 117201 (2022).

\bibitem{similar_theory} M. A. Kuznetsov and A. A. Fraerman, Temperature-sensitive spin-wave nonreciprocity induced by interlayer dipolar coupling in ferromagnet/paramagnet and ferromagnet/superconductor hybrid systems, Phys. Rev. B \textbf{105}, 214401 (2022).



\bibitem{superconductor_gate} I. A. Golovchanskiy, N. N. Abramov, V. S. Stolyarov, V. V. Bolginov, V. V. Ryazanov, A. A. Golubov, and A. V.
Ustinov,  Ferromagnet/Superconductor Hybridization for Magnonic Applications,  Adv. Funct. Mater. \textbf{28}, 1802375 (2018).

\bibitem{superconductor_gate_1} I. A. Golovchanskiy, N. N. Abramov, V. S. Stolyarov, V. V. Ryazanov, A. A. Golubov, and A. V. Ustinov, Modified dispersion law for spin waves coupled to a superconductor, J. Appl. Phys.
\textbf{124}, 233903 (2018).

\bibitem{superconductor_gate_2}  I. A. Golovchanskiy, N. N. Abramov, V. S. Stolyarov, P. S. Dzhumaev, O. V. Emelyanova, A. A. Golubov, V. V. Ryazanov, and A. V. Ustinov, Ferromagnet/Superconductor Hybrid Magnonic Metamaterials, Adv. Sci. \textbf{6}, 1900435 (2019).

\bibitem{superconductor_gate_3} I. A. Golovchanskiy, N. N. Abramov, V. S. Stolyarov, A. A. Golubov, V. V. Ryazanov, and A. V. Ustinov, Nonlinear spin waves in ferromagnetic/superconductor hybrids, J. Appl. Phys. \textbf{127}, 093903 (2020).

\bibitem{superconductor_gate_4} X. H. Zhou and T. Yu, Gating ferromagnetic resonance by superconductors via modulated reflection of magnetization-radiated electric fields. Phys. Rev. B \textbf{108}, 144405 (2023).

\bibitem{Borst} M. Borst, P. H. Vree, A. Lowther, A. Teepe, S. Kurdi, I. Bertelli, B. G. Simon, Y. M. Blanter, and T. van der Sar, Observation and control of hybridspin-wave–Meissner-current transport modes, Science \textbf{382}, 430 (2023).





 
\bibitem{Ohmic_conductors5} M. L. Sokolovskyy, J. W. Klos, S. Mamica, and M. Krawczyk, Calculation of the spin-wave spectra in planar magnonic crystals with metallic overlayers,
J. Appl. Phys. \textbf{111}, 07C515 (2012).


\bibitem{Ohmic_conductors2}  M. Mruczkiewicz and M. Krawczyk, Nonreciprocal dispersion of spin waves in ferromagnetic thin films covered with a finite-conductivity metal,
J. Appl. Phys. \textbf{115}, 113909  (2014).

\bibitem{Ohmic_conductors3}  M. Mruczkiewicz, E. S. Pavlov, S. L. Vysotsky, M. Krawczyk, Y. A. Filimonov, and S. A. Nikitov, Observation of magnonic band gaps in magnonic crystals
with nonreciprocal dispersion relation, Phys. Rev. B \textbf{90},  174416 (2014).

\bibitem{Ohmic_conductors4}  M. Mruczkiewicz, P. Graczyk, P. Lupo, A. Adeyeye, G. Gubbiotti, and M. Krawczyk, Spin-wave nonreciprocity and magnonic band structure in a
thin permalloy film induced by dynamical coupling with an array of Ni stripes, Phys. Rev. B \textbf{96}, 104411  (2017).






\bibitem{eddy_currents1}  I. Bertelli, B. G. Simon, T. Yu, J. Aarts, G. E. W. Bauer, Y. M. Blanter, and T. van der Sar, Imaging spin-wave damping underneath metals using electron spins in diamond, Adv. Quant. Technol. \textbf{4}, 2100094  (2021).

\bibitem{eddy_currents2}  P. Pincus, Excitation of spin waves in ferromagnets: eddy current and boundary condition effects, Phys. Rev. \textbf{118}, 658 (1960).

\bibitem{eddy_currents3} M. Kostylev, Strong asymmetry of microwave absorption by bilayer conducting ferromagnetic films in the microstrip-line based broadband ferromagnetic resonance, J. Appl. Phys. \textbf{106}, 043903 (2009).

\bibitem{eddy_currents4} M. A. Schoen, J. M. Shaw, H. T. Nembach, M. Weiler, and T. J. Silva, Radiative damping in waveguide-based ferromagnetic resonance measured via analysis of perpendicular standing spin waves in sputtered permalloy films, Phys. Rev. B \textbf{92}, 184417 (2015).

\bibitem{eddy_currents9}
V. Flovik, F. Maci{\' a}, A. D. Kent, and E. Wahlstr{\" o}m, Eddy current interactions in a ferromagnet-normal metal bilayer structure, and its impact on ferromagnetic resonance lineshapes, J. Appl. Phys. \textbf{117}, 143902 (2015).


\bibitem{eddy_currents10}
V. Flovik, B. H. Pettersen, and E. Wahlstr{\" o}m, Eddy-current effects on ferromagnetic resonance: Spin wave excitations and microwave screening effects, J. Appl. Phys. \textbf{119}, 163903 (2016).

\bibitem{eddy_currents5}  Y. Li and W. E. Bailey, Wave-Number-Dependent Gilbert Damping in Metallic Ferromagnets, Phys. Rev. Lett. \textbf{116}, 117602 (2016).

\bibitem{eddy_currents6}  M. Kostylev, Coupling of microwave magnetic dynamics in thin ferromagnetic films to stripline transducers in the geometry of the broadband stripline ferromagnetic resonance, J. Appl. Phys. \textbf{119}, 013901 (2016).
  
\bibitem{eddy_currents7}  J. W. Rao, S. Kaur, X. L. Fan, D. S. Xue, B. M. Yao, Y. S. Gui, and C. M. Hu, Characterization of the non-resonant radiation damping in coupled cavity photon magnon system, Appl. Phys. Lett. \textbf{110}, 262404 (2017). 


\bibitem{eddy_currents11}
S. Balaji and M. Kostylev, A two dimensional analytical model for the study of ferromagnetic resonance responses of single and multilayer films,
J. Appl. Phys. \textbf{121}, 123906 (2017).
\bibitem{eddy_currents12}
A. A. Nikitin, V. V. Vitko, A. A. Nikitin, A. B. Ustinov, V. V. Karzin, A. E. Komlev, B. A. Kalinikos, and E. L{\" a}hderanta, Spin-Wave Phase Shifters Utilizing Metal–Insulator Transition, IEEE Magn. Lett. \textbf{9}, 3706905 (2018).
\bibitem{eddy_currents13}
O. Gladii, R. L. Seeger, L. Frangou, G. Forestier, U. Ebels, S. Auffret, and V. Baltz, Stacking order-dependent sign-change of microwave phase due to eddy currents in nanometer-scale NiFe/Cu heterostructures, Appl. Phys. Lett. \textbf{115}, 032403 (2019).
\bibitem{eddy_currents14}
Y. Zhang, D. Cai, C. Zhao, M. Zhu, Y. Gao, Y. Chen, X. Liang, H. Chen, J. Wang, Y. Wei, Y. He, C. Dong, N. Sun, M. Zaeimbashi, Y. Yang, H. Zhu, B. Zhang, K. Huang, and N. X. Sun, Nonreciprocal Isolating Bandpass Filter With Enhanced Isolation Using Metallized Ferrite, IEEE Trans. Microw. Theory Tech. \textbf{68}, 5307 (2020).
\bibitem{eddy_currents15}
R. O. Serha, D. A. Bozhko, M. Agrawal, R. V. Verba, M. Kostylev, V. I. Vasyuchka, B. Hillebrands, and A. A. Serga, Low-Damping Spin-Wave Transmission in YIG/Pt-Interfaced Structures, Adv. Mater. Interfaces \textbf{9}, 2201323 (2022).


\bibitem{eddy_currents8}  S. A. Bunyaev, R. O. Serha, H. Y. Musiienko-Shmarova, A. J. Kreil, P. Frey, D. A. Bozhko, V. I. Vasyuchka, R. V. Verba, M. Kostylev, B. Hillebrands, G. N.
Kakazei, and A. A. Serga, Spin-wave relaxation by eddy currents in Y$_3$Fe$_5$O$_{12}$/Pt bilayers and a way to suppress it, Phys. Rev. A \textbf{14},  024094 (2020).



\bibitem{barrier_1} Z. R. Yan, C. H. Wan, and X. F. Han, Magnon Blocking Effect in an Antiferromagnet-Spaced Magnon Junction, Phys. Rev. Appl. \textbf{14}, 044053 (2020). 

\bibitem{barrier_2} T. Schneider, A. A. Serga, B. Leven, B. Hillebrands, R. L. Stamps, and M. P. Kostylev, Realization of spin-wave logic gates, Appl. Phys. Lett. \textbf{92}, 022505 (2008). 

\bibitem{semitransparent} M. Kostylev, Magnonic Hong-Ou-Mandel effect, Phys. Rev. B \textbf{108}, 134416 (2023).

\bibitem{barrier_4} J. Fransson, A. M. Black-Schaffer, and A. V. Balatsky, Magnon Dirac materials, Phys. Rev. B \textbf{94}, 075401 (2016).

\bibitem{barrier_3} R. J. Doornenbal, A. Rold{\'a}n-Molina, A. S. Nunez, and R. A. Duine, Spin-Wave Amplification and Lasing Driven by Inhomogeneous Spin-Transfer Torques, Phys. Rev. Lett. \textbf{122}, 037203 (2019).



\bibitem{barrier_5} S. H. Yuan, C. W. Sui, Z. D. Fan, J. Berakdar, D. S. Xue, and
C. L. Jia, Magnonic Klein and acausal tunneling enabled
by breaking the anti parity-time symmetry in
antiferromagnets, Commun. Phys. \textbf{6}, 95 (2023).

\bibitem{barrier_6} J. S. Harms, H. Y. Yuan, and R. A. Duine, Realizing the bosonic Klein paradox in a magnonic system, arXiv:2109.00865.


\bibitem{surface_roughness} T. Yu, S. Sharma, Y. M. Blanter, and G. E. W. Bauer, Surface dynamics of rough magnetic films, Phys. Rev. B \textbf{99}, 174402 (2019).

\bibitem{scratches} M. Mohseni, R. Verba, T. Br\"acher, Q. Wang, D. A. Bozhko, B. Hillebrands, and P. Pirro, Backscattering Immunity of Dipole-Exchange Magnetostatic Surface Spin Waves, Phys. Rev. Lett. \textbf{122}, 197201 (2019).



\bibitem{supplement} See Supplemental Material [...] for the details of the radiated electromagnetic fields and the non-Hermitian skin effect by chiral damping.   


\bibitem{Jackson}  J. D. Jackson, \textit{Classical Electrodynamics} (Wiley, New York, 1998).



\bibitem{Griffiths} D. J. Griffiths, \emph{Introduction to electrodynamics}, (Cambridge University Press, Cambridge, UK, 1999).

\bibitem{YIG1} J. Chen, T. Yu, C. Liu, T. Liu, M. Madami, K. Shen, J. Zhang, S. Tu, M. S. Alam, K. Xia, M. Wu, G. Gubbiotti, Y. M. Blanter, G. E. W. Bauer, and H. Yu, Excitation of unidirectional exchange spin waves by a nanoscale magnetic grating, Phys. Rev. B \textbf{100}, 104427 (2019).

\bibitem{FS_no_shift_1} L. -L. Li, Y. -L. Zhao, X. -X. Zhang, and Y. Sun, Possible evidence for spin-transfer torque induced by spin-triplet supercurrents, Chin. Phys. Lett. \textbf{35}, 077401 (2018).

\bibitem{FS_no_shift_2} K. -R. Jeon, C. Ciccarelli, H. Kurebayashi, L. F. Cohen, X. Montiel, M. Eschrig, T. Wagner, S. Komori, A. Srivastava, J. W. A. Robinson, and M. G. Blamire, Effect of Meissner Screening and Trapped Magnetic Flux on Magnetization Dynamics in Thick $\mathrm{Nb}/{\mathrm{Ni}}_{80}{\mathrm{Fe}}_{20}/\mathrm{Nb}$ Trilayers, Phys. Rev. Appl. \textbf{11}, 014061 (2019).

\bibitem{Beach} B. H. Lee, T. Fakhrul, C. A. Ross, and G. S. D. Beach, Large Anomalous Frequency Shift in Perpendicular Standing Spin Wave Modes in BiYIG Films Induced by Thin Metallic Overlayers, Phys. Rev. Lett. \textbf{130}, 126703 (2023).

\bibitem{spin_pumping} Y. Tserkovnyak, A. Brataas, and G. E. W. Bauer, Enhanced Gilbert Damping in Thin Ferromagnetic Films, Phys. Rev. Lett. \textbf{88}, 117601 (2002).

\bibitem{quantum_transport_Yuli} Y. V. Nazarov and Y. M. Blanter, \textit{Quantum Transport:
Introduction to Nanoscience} (Cambridge University Press, Cambridge, England, 2009).

\bibitem{Sakurai} J. J. Sakurai, and J. Napolitano, \emph{Modern Quantum Mechanics Second Edition}, (Cambridge University Press, Cambridge, UK, 2017).

\bibitem{skin1} K. Zhang, Z. Yang, and C. Fang, Correspondence between winding numbers and skin modes in non-Hermitian systems, Phys. Rev. Lett. \textbf{125}, 126402 (2020).
\bibitem{skin2} N. Okuma, K. Kawabata, K. Shiozaki, and M. Sato, Topological origin of non-Hermitian skin effects, Phys. Rev. Lett. \textbf{124}, 086801 (2020).
\bibitem{skin3} H. Hu and E. Zhao, Knots and non-Hermitian Bloch bands, Phys. Rev. Lett. \textbf{126}, 010401 (2021). 


\bibitem{Bergholtz} E. J. Bergholtz, J. C. Budich, and F. K. Kunst, Exceptional topology of non-Hermitian systems, Rev. Mod. Phys. \textbf{93}, 015005 (2021).



\bibitem{SAWs_1} L. Shao, D. Zhu, M. Colangelo, D. Lee, N. Sinclair, Y. Hu, P. T. Rakich, K. Lai, K. K. Berggren, and M. Lon\ifmmode \check{c}\else \v{c}\fi{}ar, Electrical control of surface acoustic waves. Nat. Electron. \textbf{5}, 348 (2022).

\bibitem{SAWs_2} L. Liao, J. Puebla, K. Yamamoto, J. Kim, S. Maekawa, Y. Hwang, Y. Ba, and Y. Otani, Valley-Selective Phonon-Magnon Scattering in Magnetoelastic Superlattices, Phys. Rev. Lett. \textbf{131}, 176701 (2023).

\bibitem{Canming} Y. C. Han, C. H. Meng, H. Pan, J. Qian, Z. J. Rao, L. P. Zhu, Y. S. Gui, C.-M Hu, and Z. H. An, Bound chiral magnonic polariton states for ideal microwave isolation, Sci. Adv. \textbf{9}, eadg4730(2023).

\bibitem{Lin} H. Xie, L.-W. He, X. Shang, G.-W. Lin, and X.-M. Lin, Nonreciprocal photon blockade in cavity optomagnonics, Phys. Rev. A \textbf{106}, 053707 (2022).

\bibitem{chiral_photonics_1} P. Lodahl, S. Mahmoodian, S. Stobbe, A. Rauschenbeutel,
P. Schneeweiss, J. Volz, H. Pichler, and P. Zoller, Chiral quantum optics, Nature (London) \textbf{541}, 473 (2017).

\bibitem{chiral_photonics_2} C. Genet, Chiral Light–Chiral Matter Interactions: an Optical Force Perspective, ACS Photon. \textbf{9}, 319 (2022).




\end{thebibliography}
\end{document}